\documentclass[prd]{revtex4}

\usepackage{amsfonts,amsmath}
\usepackage[dvips]{graphicx}

\begin{document}

\title{Dark energy and global rotation of the Universe}

\author{W{\l}odzimierz God{\l}owski}
\email{godlows@oa.uj.edu.pl}

\author{Marek Szyd{\l}owski}
\email{uoszydlo@cyf-kr.edu.pl}

\affiliation{Astronomical Observatory, Jagiellonian University, \\
Orla 171, 30-244 Krakow, Poland}

\begin{abstract}
We discuss the problem of universe acceleration driven by global rotation.
The redshift-magnitude relation is calculated and discussed in the context
of SN Ia observation data. It is shown that the dynamics of considered 
problem is equivalent to the Friedmann model with additional non-interacting 
fluid with negative pressure. We demonstrate that the universe acceleration
increase is due to the presence of global rotation effects, although
the cosmological constant is still required to explain the SN Ia data.
We discuss some observational constraints coming from SN Ia imposed on
the behaviour of the homogeneous Newtonian universe in which matter
rotates relative local gyroscopes. In the Newtonian theory 
$\Omega_{\text{r},0}$ can be identified with $\Omega_{\omega,0}$ (only 
dust fluid is admissible) and rotation can exist with 
$\Omega_{\text{r},0}=\Omega_{\omega,0} \le 0$. 
However, the best-fit flat model is the model without rotation, i.e.,
$\Omega_{\omega,0}=0$. In the considered case we obtain the limit for 
$\Omega_{\omega,0}>-0.033$ at $1\sigma$ level. We are also beyond the model 
and postulate the existence of additional matter which scales like radiation 
matter and then analyse how that model fits the SN Ia data. In this case the 
limits on rotation coming from BBN and CMB anisotropies are also obtained.
If we assume that the current estimates are $\Omega_{\text{m},0}\sim 0.3$, 
$\Omega_{\text{r},0} \sim 10^{-4}$, then the SN Ia data show that 
$\Omega_{\omega,0} \ge -0.01$ 
(or $\omega_0 < 2.6 \cdot 10^{-19}\, \text{rad}/\text{s}$).
The statistical analysis gives us that the interval for any matter scaling
like radiation is $\Omega_{\text{r},0} \in (-0.01, 0.04)$.
\end{abstract}

\maketitle

\section{Introduction}
 
We consider a homogeneous universe which is more general than the FRW
model, in which matter additionally rotates relative to localgyroscopes 
\cite{Godlowski03}. The motion of the fluid in such a universe is 
described by the scalar expansion $\theta$, the rotation tensor 
$\omega_{ab}$, and the shear tensor $\sigma_{ab}$. The homogeneous 
rotation of fluid as a whole is usually called the global rotation of 
the universe \cite{Li98}.
 
Applying these concepts we must remember that CMB strongly restricts
(indirectly from observations) the value of angular velocity
\cite{Collins73,Hawking74}. On the other hand one can estimate
the present value of $\theta_0$, $\omega_0$, and $\sigma_0$ also
directly from observations of galaxies \cite{Kristian66}. These
observations show that $\theta_0 = 3H_0$, $\omega_0 \lesssim \theta_0/3$,
$\sigma_0 \lesssim \theta_0/4$.
 
The propagation equation for $\theta$, known as the Raychaudhuri equation
\cite{Ellis73,Ciufolini95}, for the perfect fluid with energy-momentum tensor
$T_{ab} = (\rho + p)u_a u_b + pg_{ab}$ (where $\rho$ and $p$ are the energy
density and pressure, respectively), is
\begin{equation}
\label{eq:1}
\dot{\Theta} - \dot{u}_{; a}^{a} + \frac{1}{3} \Theta^2
+ 2(\sigma^2 - \omega^2) + \frac{1}{2}(\rho + 3p) - \Lambda = 0
\end{equation}
where $\dot{u}_{a} \equiv u_{a;b}u^{b}$ is the acceleration vector; we 
shall use a dot to denote the rate of change of any quantity as measured 
by an observer moving with $4$-velocity $u^a$; and $\omega^2 = \omega_{ab} 
\omega^{ab}/2$, $\sigma^2 = \sigma_{ab} \sigma^{ab}/2$ are the scalars of 
rotation and shear, respectively; and $\Lambda $ is the cosmological 
constant. 
 
If we define a representative length $l$ along a particle world line by
\cite{Ellis73,Ciufolini95}
\begin{equation}
\label{eq:2}
\frac{\dot{l}}{l} = \frac{1}{3}\Theta
\end{equation}
then $l$ represents the volume behaviour of the fluid completely. For 
example from $l$ one can define the Hubble function $H$ and the deceleration
parameter $q$ by
\begin{equation}
\label{eq:3}
H \equiv \frac{\dot{l}}{l}, \qquad
q \equiv - \frac{\ddot{l}}{l} H.
\end{equation}
Using definition (\ref{eq:2}) and (\ref{eq:3}), equation (\ref{eq:1})
can be rewritten in the form
\begin{equation}
\label{eq:4}
3 \frac{\ddot{l}}{l} = 2(\omega^2 - \sigma^2) + \dot{u}_{\phantom{a};a}^{a}
- \frac{1}{2}(\rho + 3p) + \Lambda.
\end{equation}
This shows how the acceleration of the universe (the curvature of
curve $l(t)$) is directly determined at each point of spacetime.
Let us note that $\Lambda$ acts as a constant repulsive force
whereas rotation as a variable repulsive force.
 
When $\omega^2$, $\sigma^2$, and $\dot{u}_{\phantom{a};a}^{a}$ are given as
a function of $l$ we can integrate equation (\ref{eq:4}).
To simplify matter we take $\dot{u}_{a} = 0$ (because the acceleration
vector represents the effects of non-gravitational forces it vanishes
when a particle moves along a geodesic, which would necessarily follow
in the case of dust).
 
It has been shown that spatially homogeneous, rotating, and expanding
universes with the perfect fluid have the non-vanishing shear
\cite{Ellis73,Ciufolini95}. This is quite contrary to the case of
the homogeneous Newtonian cosmology where many such solutions are known.
These homogeneous shear-free solutions are independent of the pressure
which may be set equal to zero or a constant. This difference in
the two theories seems to be both surprising and interesting since
Ellis' theorem has a purely local character, and it is completely
independent of the strength of the gravitational field
\cite{Ellis73,Ciufolini95,Senovilla98}.
 
Let us consider solutions with $\sigma = 0 = \dot{u}$, $\omega \Theta = 0$.
In this case $\omega^2 = \omega_0^2 / l^4$ where $\dot{\omega}=0$. Then
we can integrate the Raychaudhuri equation using the conservation
equation
\begin{equation}
\label{eq:5}
\dot{\rho} + \Theta(\rho + p) = 0.
\end{equation}
The occurrence of term $p$ in the factor $(\rho + p)$ is a special
relativistic effect \cite{Ellis73,Ciufolini95}.
 
In the considered case of $\omega \Theta \ne 0$ we obtain the generalised
Friedmann equation
\begin{subequations}
\label{eq:6}
\begin{align}
3 \dot{l}^2 + 2 \frac{\omega_0^2}{l^2} - \frac{\mu l^3}{l}
- \Lambda l^2 &= - 3k \\
\dot{k} &= 0
\end{align}
\end{subequations}
where $\mu = \text{const}$, $p=0$ and $l(t) = a(t)$ is the scale factor.
From the mathematical point of view equation (\ref{eq:6}) is a first
integral of system (\ref{eq:4}).
 
Equation (\ref{eq:6}) can be treated as basic equations in a Newtonian
homogeneous cosmology. Solutions of this equation represent shear-free
Newtonian cosmologies which are in general both expanding $(i \ne 0)$
and rotating $(\omega_{0} \ne 0)$. Equation (\ref{eq:6}) is called
the Heckmann-Sch\"uking equation \cite{Heckmann59}.
 
If we consider models with $\omega = 0 = \dot{u}$, $\sigma \Theta \ne 0$,
and the Ricci tensor  ${}^3 R_{ab}$ is isotropic then we obtain $\sigma^2 =
\Sigma^2/l^6$ where $\Sigma =0$. We can then integrate the Raychaudhuri
equation to obtain the generalised Friedmann equation
\begin{subequations}
\label{eq:7}
\begin{align}
3 \dot{l}^2 - \frac{\Sigma^2}{l^4} - \frac{\mu l^3}{l}
- \Lambda l^2 &= - 3k \\
\dot{k} &= 0
\end{align}
\end{subequations}
where $l = (a_1 a_2 a_3)^{1/3}$ is an average scale factor.
 
Therefore, it seems reasonable to assume that $\sigma$ is sufficiently
small compared with $\omega$ since the shear falls off more rapidly
than the rotation \cite{Li98,Ellis73,Ciufolini95}.
 
In the case of dust $\sigma^2 \propto a^{-6}$ whereas
$\omega^2 \propto a^{-4}$. The conservation of angular momentum gives
$\omega \rho a^5 = \text{const}$ \cite{Ellis73,Ciufolini95}.
 
From equation (\ref{eq:7}) we see that the effect of anisotropy is like 
in the FRW model with stiff matter. In our further analysis of observational 
effects we consider equation~(\ref{eq:6}) as a simplest model in which the 
effect of global rotation can be investigated. However, we also consider 
the presence of additional non-interacting radiation matter which can be 
treated as a simple extension beyond the Newtonian model.

\section{Effect of global rotation on acceleration of the universe}
 
The supernovae observations indicate that the Universe's expansion has
started to accelerate during recent cosmological times, and CMB observations
suggest that the Universe is dominated by a dark energy component, with
negative pressure, driving the acceleration \cite{Perlmutter99,Riess98}.
While the most obvious candidate for such a component is the vacuum energy
a plausible alternative is the dynamical vacuum energy or quintessence.
However, these models usually face fine-tuning problems, because there
is a question of explaining why the vacuum energy dominates the Universe
only recently \cite{Caldwell98}.
 
To study the effect of the global rotation on the acceleration of
the universe we formally introduce rotation to the model by definition of
\begin{subequations}
\label{eq:8}
\begin{align}
\rho_{\text{eff}} &= \rho_{\text{m}} + \rho_{\omega}
= \rho_{\text{m} 0} a^{-3} + \rho_{\omega 0} a^{-4} + \Lambda \\
p_{\text{eff}} &= \frac{1}{3} \rho_{\omega} - \Lambda
\end{align}
\end{subequations}
where $\rho_{\omega 0} = -2 \omega_0^{2} < 0$ and $p_{\omega} =
\frac{1}{3}\rho_{\omega}$ (like for radiation matter).
 
Therefore, in the case of dust filled universe, the dynamical effect of 
global rotation is equivalent to an additional non-interacting fluid with 
negative pressure.
 
In order to take into account the effects of rotation we introduce
\begin{subequations}
\label{eq:9}
\begin{align}
\Omega_{\omega} &= \frac{\rho_{\omega}}{3H_{0}^{2}}
= - \frac{2\omega_0^{2}}{3H_{0}^{2}}
\left( \frac{a}{a_0} \right)^{-4} \\
\Omega_{\text{m}} &= \frac{\rho_{\text{m}}}{3H_{0}^{2}}
= \frac{\rho_{0\text{m}}}{3H_{0}^{2}}
\left( \frac{a}{a_0} \right)^{-3}.
\end{align}
\end{subequations}
 
For our purpose it is also useful to rewrite the dynamical equations to
a new form using dimensionless quantities
\[
x \equiv \frac{a}{a_{0}}, \qquad T \equiv |H_{0}| t
\]
with $H=\dot{a}/a$, $\rho_{\text{cr},0} \equiv 3H_{0}^{2}$ and the
subscript $0$ means that a quantity with this subscript is evaluated
today (at time $t_0$). Additionally we define $\Omega_{k,0} =
- 3k/6H_{0}^{2}$ and $\Omega_{\Lambda,0} = \Lambda/3H_{0}^{2}$.
 
The basic dynamical equations are then rewritten as \cite{Perlmutter99}
\begin{subequations}
\label{eq:10}
\begin{align}
\frac{\dot{x}^{2}}{2} &= \frac{1}{2} \Omega_{k,0} + \sum_{i}
\Omega_{i,0} x^{-1-3\gamma_{i}} \\
\ddot{x} &= - \frac{1}{2} \sum_{i} \Omega_{i,0}(1+3\gamma_{i})
x^{-2-3\gamma_{i}}
\end{align}
\end{subequations}
where $i=(\text{m},\omega,\Lambda)$. The above equations can be represented
as the two-dimensional dynamical system
\begin{subequations}
\label{eq:11}
\begin{align}
\dot{x} &= y \\
\dot{y} &= -\frac{1}{2} \sum_{i} \Omega_{i,0} (1+3\gamma_{i})x^{-2-3w_{i}}
\end{align}
\end{subequations}
or by the Hamiltonian dynamical system with the Hamiltonian given in the form
\begin{equation}
\label{eq:12}
\mathcal{H} = \frac{1}{2} \dot{x}^{2} + V(x) \equiv 0
\end{equation}
and with the potential
\[
V(x) = - \frac{1}{2}\Omega_{k,0} - \sum_{i} \Omega_{i,0} x^{-1-3\gamma_{i}}.
\]
The system should be considered on the zero energy level.
 
The form of (\ref{eq:12}) can be useful in particle-like description
for the simplest model with global rotation, whereas form (\ref{eq:11})
is helpful in the analysis of dynamics on a phase plane $(x,y)$.
 
The system under consideration can be identified after taking
\begin{align*}
w_1 &= 1/3 & & (\text{effect of rotation or radiation}) \\
w_2 &= 0   & & (\text{effect of dust matter}) \\
w_3 &= -1  & & (\text{effect of } \Lambda,\ \rho=\Lambda)
\end{align*}
As an example of application of these equations consider the case of
$\Omega_{\text{m},0}, \Omega_{\omega,0}, \Omega_{\Lambda,0} \ne 0$.
Then our Universe accelerates provided that the potential $V$ is a 
decreasing function of its argument 
\begin{equation}
-\frac{dV}{dx} = - \Omega_{\omega,0}x^{-3} - \frac{1}{2} \Omega_{\text{m},0}
x^{-2} + \Omega_{\Lambda,0} x > 0,
\end{equation}
i.e., if $\Omega_{\text{m},0}=0$, the universe always accelerates for 
every $x$.
 
For $\Omega_{\Lambda,0}=0$ the Universe accelerates provided that
\[
x < - \frac{\Omega_{\omega,0}}{\Omega_{\text{m},0}}
\]
where $\Omega_{\omega,0} < 0$ and $\sum_{i} \Omega_{i,0}+\Omega_{k,0}=1$
($i=\text{m},\omega,\Lambda$).

Our Universe accelerates at present provided that
\[
-2 - \Omega_{\text{m},0} + 4 \Omega_{\Lambda,0} + 2 \Omega_{k,0} >0
\]
or $\Omega_{\Lambda,0}>0.425$ where we assume $\Omega_{k,0}=0$,
$\Omega_{\text{m},0}=0.3$. We can see that rotation lowers the value of
cosmological constant needed to explain the SN Ia data.

\section{Magnitude-redshift relation in the model}
 
The important test to verify whether rotation may represent ``dark energy''
(which can be called true dark radiation because causes the acceleration of
the Universe) is to compare rotation effects with the supernovae type Ia 
data. The answer is that global rotation may be seriously taken as 
a candidate to describe only part of dark energy and the cosmological 
constant is still required.

It is well known that cosmic distance measures, like the luminosity distance,
depend sensitively on the spatial geometry (curvature) and dynamics. Therefore,
luminosity depends on the present densities of the different components
of matter content and their equations of state. For this reason, the
magnitude-redshift relation for distant objects is proposed as a potential
test for cosmological models and play an important role in determining
cosmological parameters.

Let us consider an observer located at $r=0$ at the moment $t=t_0$ who
receives light emitted at $t=t_1$ from a source of absolute luminosity $L$
located at the radial distance $r_1$. Of course the cosmological redshift 
$z$ of the source is related with $t_1$ and $t_0$ by the relation 
$1+z=a(t_0)/a(t_1)$. If the apparent luminosity of the source measured by 
the observer is $l$, the luminosity distance $d_L$ of the source, defined by
\begin{equation}
\label{eq:59}
l = \frac{L}{4\pi d_L^2}
\end{equation}
is
\begin{equation}
\label{eq:60}
d_L = (1+z)a_0 r_1.
\end{equation}
 
For historical reasons, the observed and absolute luminosities are
defined in terms of K-corrected observed and absolute magnitudes $m$ and $M$, 
respectively ($l=10^{-2m/5}\cdot 2.52 \cdot 10^{-5}\,\text{erg}\, 
\text{cm}^{-2}\,\text{s}^{-2}$, $L=10^{-2M/5}\cdot 3.02 \cdot
10^{35}\,\text{erg}\,\text{s}^{-2}$) \cite{Weinberg72}.
When written in terms of $m$ and $M$, equation~(\ref{eq:59}) yields
\begin{equation}
\label{eq:61}
m(z,{\cal M},\Omega_{\text{m},0},\Omega_{\Lambda,0}) = {\cal M}
+ 5\log_{10}[{\cal D}_L(z,\Omega_{\text{m},0},\Omega_{\Lambda,0})]
\end{equation}
where
\begin{equation}
\label{eq:61a}
{\cal M}=M-5\log_{10}H_0+25
\end{equation}
and
\[
{\cal D}_L(z,\Omega_{\text{m},0},\Omega_{\Lambda,0}) \equiv
H_0 d_L(z,\Omega_{\text{m},0},\Omega_{\Lambda,0},H_0)
\]
is the dimensionless luminosity distance in Mpc.
 
By using expression for the FRW spacetime metric we obtain coordinate
distance $r_1$, appearing in (\ref{eq:60})
\begin{equation}
\label{eq:62}
\psi(r_1)=\int\limits_{a_0/(1+z)}^{a_0} \frac{da}{a\dot{a}}
= -\int\limits_r^0 \frac{dr}{\sqrt{1-kr^2}},
\end{equation}
with
\begin{align}
\psi(r_1) &= \sin^{-1} r_1&  \text{for}& & k&=+1 \nonumber \\
\psi(r_1) &= r_1&            \text{for}& & k&=0          \label{eq:63} \\
\psi(r_1) &= \sinh^{-1} r_1& \text{for}& & k&=-1. \nonumber
\end{align}
By using Hamiltonian constraint (\ref{eq:12}) for the model with
dust matter, cosmological constant, curvature, and global rotation 
we obtain 
\begin{equation}
\label{eq:64}
\psi(r_1)=\frac{1}{a_0 H_0}\int\limits_0^z
[\Omega_{k,0}(1+z')^2+\Omega_{\text{m},0}(1+z')^3
+\Omega_{\omega,0}(1+z')^4+\Omega_{\Lambda,0}\}^{-1/2}dz'.
\end{equation}
We obtain finally
\begin{multline}
\label{eq:65}
{\cal D}_L((z,\Omega_{\text{m},0},\Omega_{\Lambda,0},\Omega_{\omega,0})
= \frac{(1+z)}{\sqrt{{\cal K}}}\xi\left(\sqrt{{\cal K}}\int\limits_0^z
[(1-\Omega_{\text{m},0}-\Omega_{\omega,0}-\Omega_{\Lambda,0})(1+z')^2 \right. \\
\left. \phantom{\int\limits_0^z}
+ \Omega_{\text{m},0}(1+z')^3 + \Omega_{\omega,0}(1+z')^4
+ \Omega_{\Lambda,0}]^{-1/2}dz' \right),
\end{multline}
where
\begin{align}
\xi(x) &= \sin x& \text{with}& &{\cal K}&=-\Omega_{k,0}&
\text{when}& &\Omega_{k,0}&<0 \nonumber \\
\xi(x) &= x& \text{with}& &{\cal K}&=1&
\text{when}& &\Omega_{k,0}&=0
\label{eq:66} \\
\xi(x) &= \sinh x& \text{with}& &{\cal K}&=\Omega_{k,0}&
\text{when}& &\Omega_{k,0}&>0 \nonumber
\end{align}
and
\[
\Omega_{k,0} = - \frac{k}{\dot{a}_0^2}.
\]
Thus for given ${\cal M}$, $\Omega_{\text{m},0}$, $\Omega_{\Lambda,0}$,
$\Omega_{k,0}$, $\Omega_{\omega,0}$, equations (\ref{eq:61}) and
(\ref{eq:65}) give the predicted value of $m(z)$ at a given $z$.

\section{Magnitude-redshift relation in the model --- results}
 
We decided to test our model using the Perlmutter sample \cite{Perlmutter99}.
To avoid any possible selection effect we choose the full sample
without excluding any supernova from that sample. It means that our 
basic sample is Perlmutter sample A.
We test our model using the likelihood method \cite{Riess98}.
 
Firstly, we should estimate the value of ${\cal M}$ (equation~(\ref{eq:61a}))
from the full sample of 60 supernovae taking $\Omega_{\omega,0}=0$ (the pure
Perlmutter \& Riess model). We obtain value of ${\cal M}=-3.39$
(we also assume that the present value of the Hubble constant is
$H_0=65 \,\text{km}\,\text{s}^{-1}\,\text{Mpc}^{-1}$)
what is in very good agreement with the result of Efstathiou {\it et al.\/}
\cite{Efstathiou99} and Vishwakarma \cite{Vishwakarma01} (Vishwakarma
obtains ${\cal M}_c=24.03$ for $c=1$, i.e., ${\cal M}=-3.365$). Also the
value of $\chi^2$ obtain for Perlmutter's flat model is $96.5$ what is in very
good agreement with Perlmutter's results (see Table~3 in \cite{Perlmutter99}).
(Some marginal differences are probably because in our analysis we include
both errors in measurements of magnitude and radial distances).
 
\begin{figure}
\includegraphics[width=0.75\textwidth]{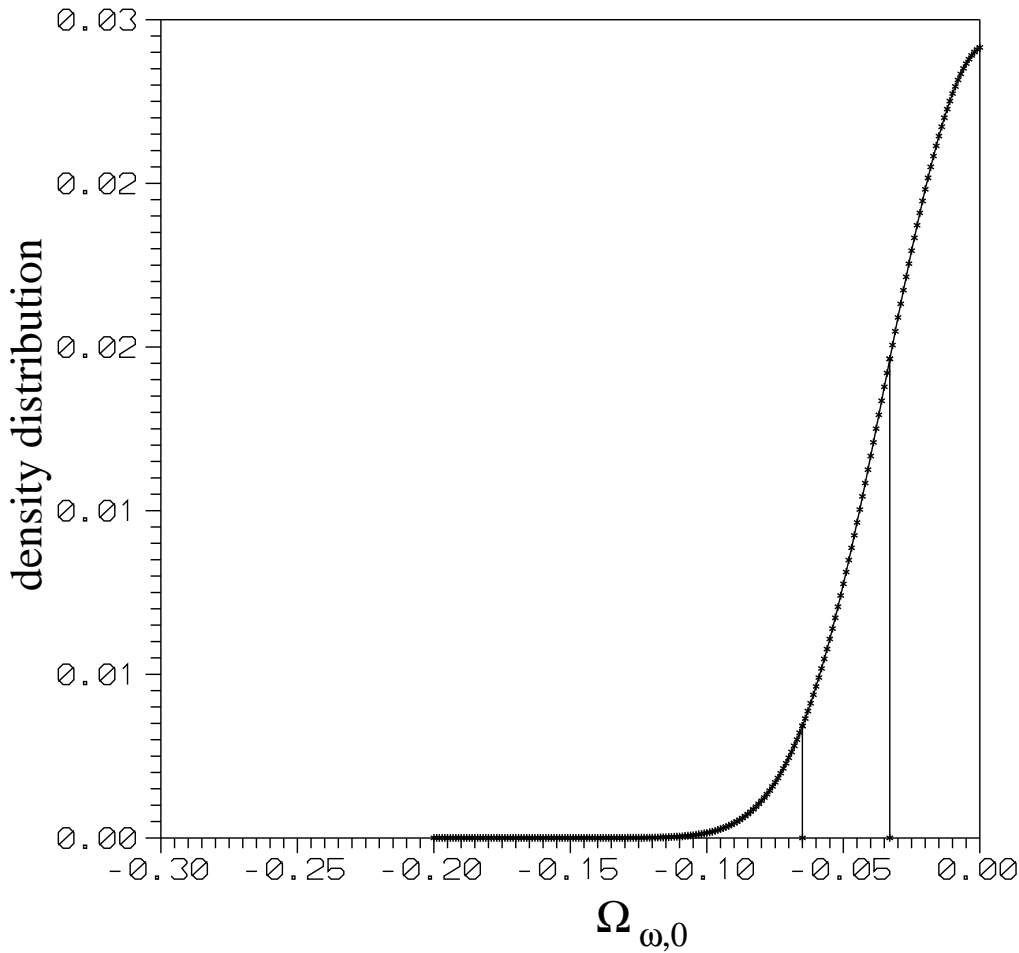}
\caption{
The density distribution for  $\Omega_{\omega,0}$ in the Newtonian model.
We obtain the limit $\Omega_{\omega,0}>-0.03$ on the $1\sigma$ level, while
$\Omega_{\omega,0}>-0.06$ on the $1\sigma$ level.}
\label{fig:brzegpo}
\end{figure}

We consider the pure Newtonian model with $\Omega_{\omega,0} < 0$ and 
assume that $\Omega_{\text{m},0} \sim 0.3$ \cite{Peebles02,Lahav02}.
Using the minimalization procedure, described below, with aforementioned
assumptions we obtain the density distribution for $\Omega_{\omega,0}$.
The results are presented on Fig.~\ref{fig:brzegpo}. Here we find that
the limit for $\Omega_{\omega,0}>-0.033$ on the $1\sigma$ level, while
$\Omega_{\omega,0}>-0.065$ on the $2\sigma$ level.

\begin{figure}
\includegraphics[width=0.75\textwidth]{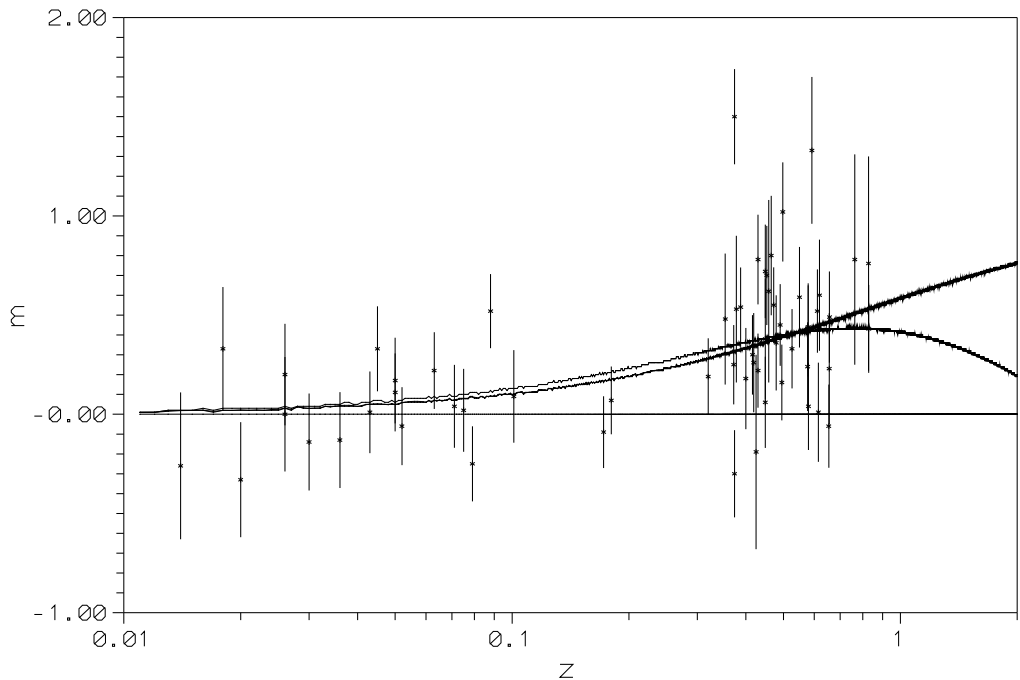}
\caption{Residuals between the Enstein-de Sitter model and three cases:
the Einstein-de Sitter itself (zero line) the best-fitted model with 
$\Omega_{\omega,0}=-0.01$ (middle curve), and the best-fitted flat model 
with $\Omega_{\omega,0}=-0.01$ and the Perlmutter model (these two curves 
overlap) (highest curve). Therefore in the case of dust matter, the 
difference between the Perlmutter model and best fit model with rotation 
becomes detectable for redshifts $z \gtrsim 1.2$.}
\label{fig:rezuj1}
\end{figure}

The analysis of the pure Newtonian model is presented on Fig.~\ref{fig:rezuj1} 
with the magnitude-redshift relation for real data (marked with asterisks) 
and for predicted values by models. The top line is the pure Perlmutter 
flat model with $\Omega_{\text{m},0}=0.28$,
$\Omega_{\Lambda,0}=0.72$. The bottom line is the pure flat model with the 
cosmological constant $\Omega_{\Lambda,0}=0$. Between these models there are
located our models with $\Omega_{\omega,0} = -0.01$ best-fitted model (lower 
curve) and best-fitted flat model (upper curve). The latter model curve 
overlaps the Perlmutter model curve. One could observe that the difference 
between our lower best-fitted model and the Einstein-de Sitter model with 
$\Omega_{\Lambda,0}=0$ is the largest for $z$ between $0.6$ and $0.7$ and 
significantly decreases for higher redshifts. There are significant differences between predictions of
these models and Perlmutter's one where differences to the pure flat model
increase for higher redshifts. It gives us possibility to discriminate 
between the Perlmutter model and our model when data from supernovae more 
distant than $z \sim 1$ could be available. It is very important because for 
present data our model is only marginally better than the Perlmutter model.

\begin{figure}
\includegraphics[width=0.75\textwidth]{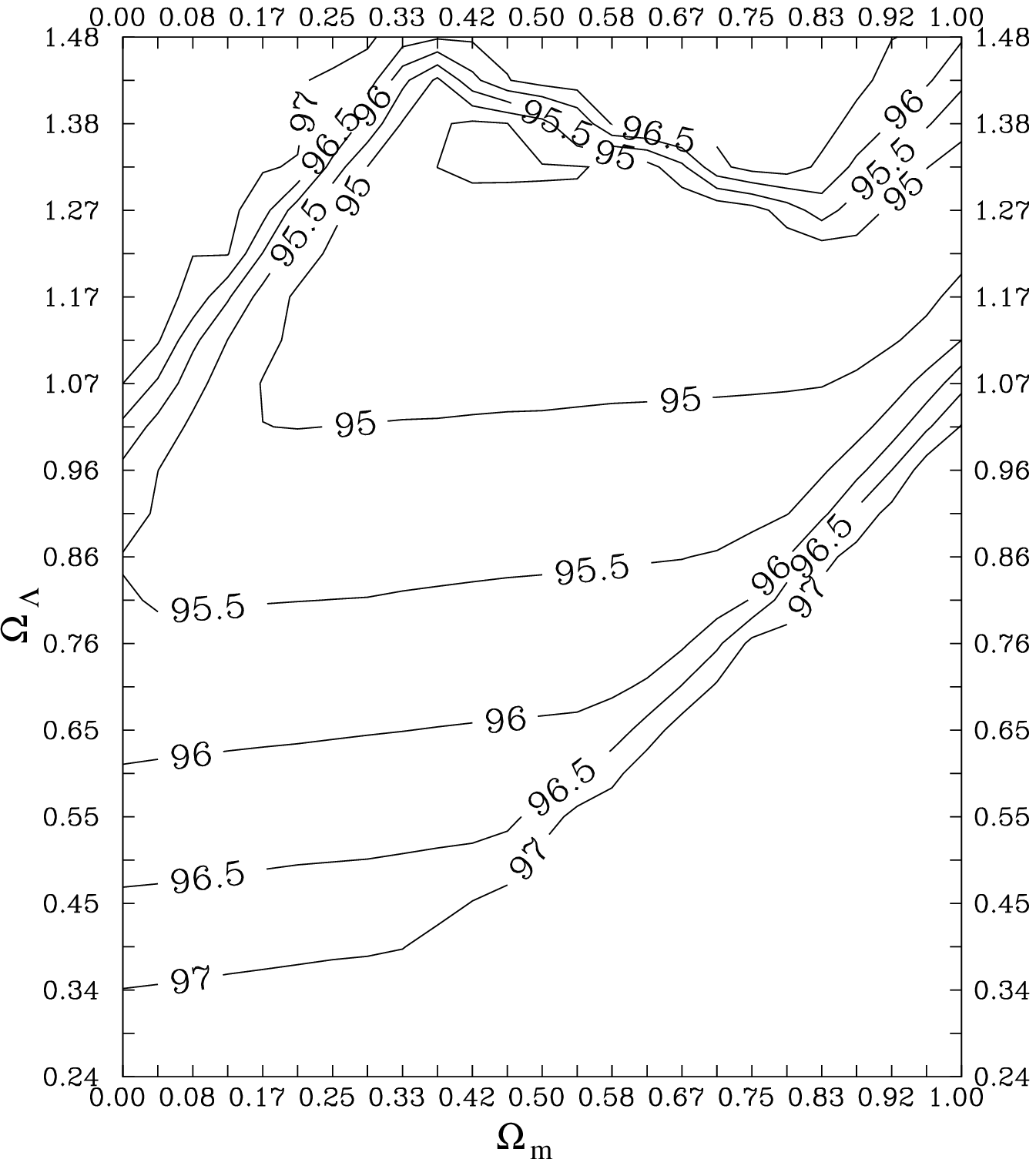}
\caption{Levels of constant $\chi^{2}$ on the plane
($\Omega_{\text{m},0},\Omega_{\Lambda,0}$) minimalized over the rest of
parameters. The figure shows the preferred value of
$\Omega_{\Lambda,0},\Omega_{\text{m},0}$.}
\label{fig:margr}
\end{figure}

We can also admit that the total matter content scales like radiation. 
It means that the contribution coming from $\Omega_{\omega,0}$ is 
included in $\Omega_{\text{r},0}$. Therefore, in the more detailed analysis 
we assumed that $\Omega_{k,0} \in [-1,1]$, $\Omega_{\text{m},0} \in [0,1]$.
From the formal point of view then we obtain the best fit
($\chi^2=94.7$) for $\Omega_{k,0}=-1.0$, $\Omega_{\text{m},0}=0.54$, 
$\Omega_{\omega,0}=0.15$, $\Omega_{\Lambda,0}=1.31$,
which is completely unrealistic. However, we should note that we obtain,
in fact, a three-dimensional ellipsoid of possible models depending on
$\Omega_{\text{m},0}$, $\Omega_{\omega,0}$, $\Omega_{\Lambda,0}$. It is more
complicated than in the case of Perlmutter's analysis when he obtains only
two-dimensional ellipsoid (depends only on $\Omega_{\text{m},0}$ and
$\Omega_{\Lambda,0}$). But, knowing the best-fit values has no enough 
scientific relevance, if not also confidence levels for parameter intervals 
are presented. On the Fig.~\ref{fig:margr} we show the levels of constant 
$\chi^{2}$ on the plane ($\Omega_{\Lambda,0},\Omega_{\text{m},0}$) minimalized 
over the rest of parameters. The figure shows the preferred value of
$\Omega_{\Lambda,0},\Omega_{\text{m},0}$. The minimalization procedure confirms
the chosen value of $\mathcal{M}=-3.39$, because it is a best-fitted value
for the flat models. Now, we would like to obtain confidence contours in the
$\Omega_{\Lambda,0},\Omega_{\text{m},0}$ plane. 

\begin{figure}
\includegraphics[width=0.75\textwidth]{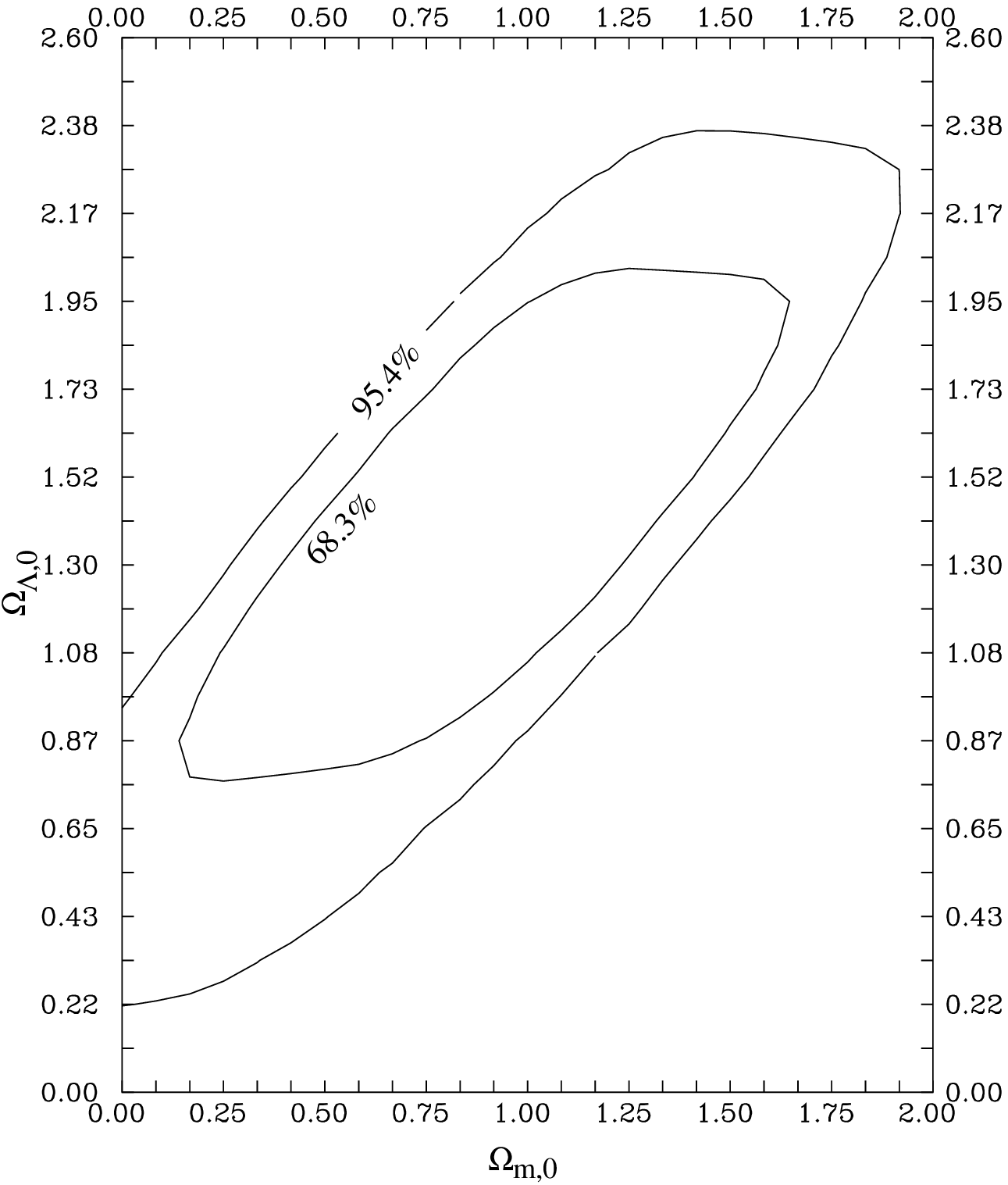}
\caption{Confidence levels on the plane
($\Omega_{\text{m},0},\Omega_{\Lambda,0}$) minimalized over the rest of
parameters for the flat model. The figure shows the ellipsoid of
the preferred value of $\Omega_{\text{m},0},\Omega_{\Lambda,0}$.}
\label{fig:max1}
\end{figure}

Since from the formal point of view we have no a priori constraints on 
cosmological parameters we assume here that $\Omega_{k,0}$ and 
$\Omega_{\text{m},0}$ are of any value. The result of our analysis are 
presented on the Fig.~\ref{fig:max1}. This figure shows the confidence 
levels of 2-dimensional distribution of $(\Omega_{\text{m},0}, 
\Omega_{\Lambda,0})$. It is analogous to the confidence level figure 
obtained by Perlmutter. 

\begin{figure}
\includegraphics[width=0.75\textwidth]{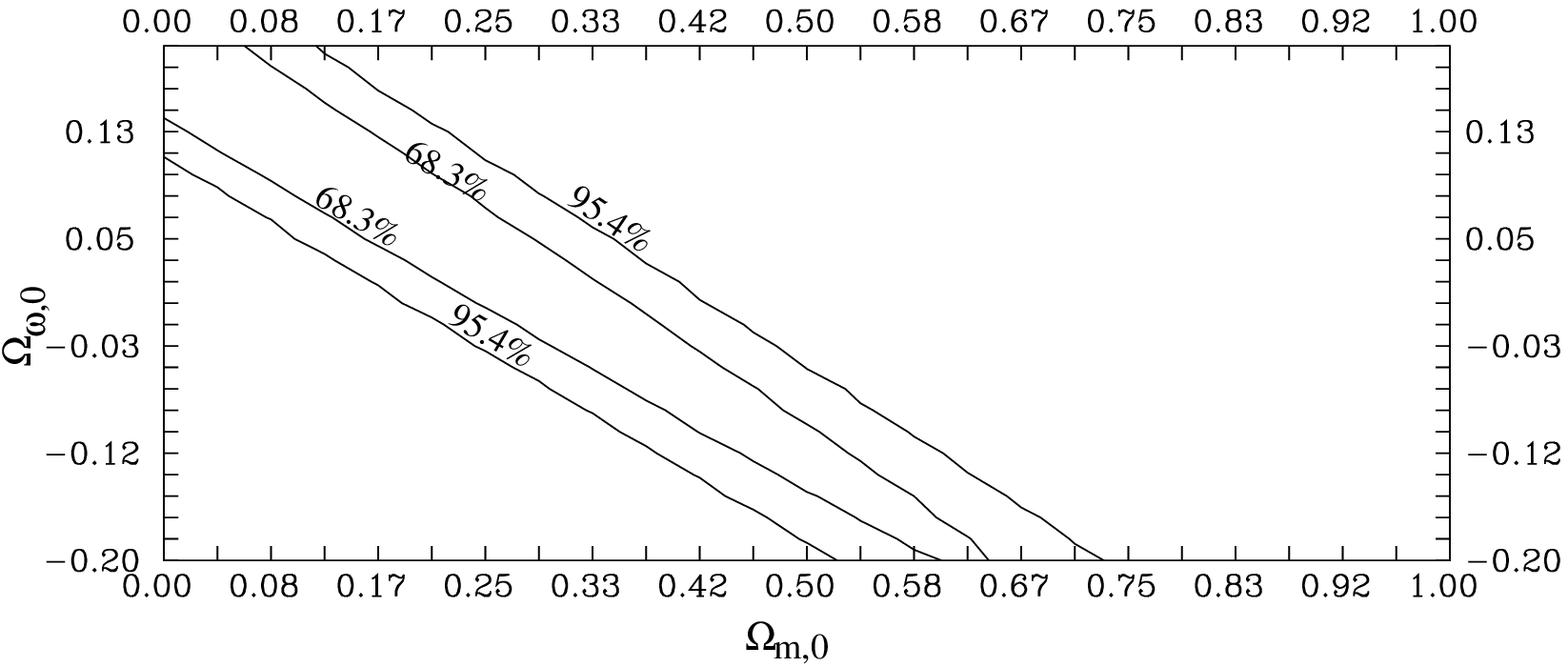}
\caption{Confidence levels on the plane
($\Omega_{\text{m,0}},\Omega_{\omega,0}$) minimalized over the rest of
parameters for the flat model. The figure shows the ellipsoid of
the preferred value of $\Omega_{\text{m,0}},\Omega_{\omega,0}$.
The results prefer the positive value of $\Omega_{\omega,0}$, while
the negative values are allowed (i.e., rotation can exist).}
\label{fig:figpl1}
\end{figure}

Another considered case is the flat model ($\Omega_{k,0}=0$) where we obtain 
``corridors'' of possible models (we presented confidence contours in 
$\Omega_{\Lambda},\Omega_{\text{m}}$ plane Fig.~\ref{fig:figpl1}). The formal 
best-fitted flat model is $\Omega_{\text{m},0}=0.12$,
$\Omega_{\omega,0}=0.12$, $\Omega_{\Lambda,0}=0.76$, $\chi^2=95.7$.
In probably a more realistic case we obtain for flat model 
$\Omega_{\text{m},0}=0.28$, $\Omega_{\omega,0}=0.02$, i.e., 
$\Omega_{\Lambda,0}=0.73$.
For that model $\chi^2=95.9$. For the flat model with low rotation 
$\Omega_{\text{m},0}=0.33$, $\Omega_{\omega,0}=-0.01$, i.e., 
$\Omega_{\Lambda,0}=0.68$, $\chi^2=96.0$. The value of $\chi^2$ is practically
the same in all three cases. It clearly shows that statistical
analysis is not sufficient for discrimination between statistically 
available models. To choose the physically plausible model we need aditional 
information which can be obtained for example from extragalactic astronomy 
investigations (especially estimations for $\Omega_{\text{m},0}$ and
$\Omega_{k,0}$ are useful).

It is interesting to observe how presence of non-zero (but rather small)
$\Omega_{k,0}$ with realistic rotation changes our results. For example
for $\Omega_{k,0}=0.1$, $\Omega_{\text{m},0}=0.28$, 
$\Omega_{\omega,0}=-0.01$, i.e. $\Omega_{\Lambda,0}=0.63$
$\chi^2=96.2$, while for $\Omega_{k,0}=-0.1$,  $\Omega_{\text{m},0}=0.38$,
$\Omega_{\omega,0}=-0.01$, i.e. $\Omega_{\Lambda,0}=0.63$,
$\chi^2=95.8$.
It shows interesting possibility if separately we could find value for
rotation $\Omega_{\omega,0}$ and matter $\Omega_{\text{m},0}$,
than we could test the value of $\Omega_{k,0}$ more precisely than with the
models without rotation.
 
From Ref.~\cite{Peebles02,Lahav02} we obtain that value $\Omega_{m,0}$
should be not far from $0.3$. With this assumption we could find from
Fig.~\ref{fig:figpl1} that $\Omega_{\omega,0}$ should satisfy
$\Omega_{\omega,0}>-0.01$ which gives critical angular velocity
$\omega_0 =2.6\cdot 10^{-19} \,\mathrm{rad}/\mathrm{s}$, is in a good agreement
with other limits, however it should be pointed that our limit is weaker,
Li \cite{Li98} suggested  $\omega_0 =6\cdot 10^{-21} \,\mathrm{rad}/\mathrm{s}$.
In terms of density parameter that limit requires
$\Omega_{\omega,0}>-5.3\cdot 10^{-6}$ whereas to obtain Ciufolini and Wheeler's 
limit \cite{Ciufolini95} is required $\Omega_{\omega,0}>-1.4\cdot 10^{-4}$.

One should note that we give our analysis without excluding any supernovae
from Perlmutter's data. However, from formal point of view, when we analyse
full Perlmutter's sample A, all analysed models should be rejected even
on the confidence level $0.99$. One of the reasons could be the fact that
assumed errors of measurements are too low. Nevertheless, another solution 
is usually suggested. We can exclude 2 supernovae as
outliers and 2 as likely reddened ones from the sample of 42 high redshift
supernovae and eventually 2 outliers from the sample of 18 low redshift
supernovae (Perlmutters's sample B and C, respectively). We decided to use 
full Perlmuttler's sample A as our basic sample because rejecting any 
supernovae from the sample could be the source of not fully controlled 
selection effect. On the other side such procedure also could be useful. 
It is the reason that we decided to check our analysis using Perlmutter's 
samples B and C. It does not significantly change our result, but 
increases quality of the fit. The formal best-fit for sample B 
(56 supernovae) is ($\chi^2=57.5$) what gives $\Omega_{k,0}=-0.3$,
$\Omega_{\text{m},0}=0.2$, $\Omega_{\omega,0}=0.17$, i.e., 
$\Omega_{\Lambda,0}=0.93$.
For the flat model we obtain ($\chi^2=57.6$)
$\Omega_{\text{m},0}=0.03$, $\Omega_{\omega,0}=0.19$, i.e., 
$\Omega_{\Lambda,0}=0.78$,
while for ``realistic'' model ($\Omega_{\text{m},0}=0.28$, $\Omega_{\omega,0}=0.03$)
$\Omega_{\Lambda,0}=0.69$  $\chi^2=57.7$.
For the flat model with small rotation $\Omega_{\omega,0}=-0.01$,
$\Omega_{\text{m},0}=0.34$, i.e., $\Omega_{\Lambda,0}=0.67$,
$\chi^2=57.8$.

The formal best-fit for sample C (54 supernovae)
($\chi^2=53.6$) gives $\Omega_{k,0}=-0.1$, $\Omega_{\text{m},0}=0.11$,
$\Omega_{\omega,0}=0.18$, i.e., $\Omega_{\Lambda,0}=0.81$,
while for flat model
$\Omega_{\text{m},0}=0.05$, $\Omega_{\omega,0}=0.19$, i.e.,
$\Omega_{\Lambda,0}=0.86$, $\chi^2=53.6$,
while for ``realistic'' model
$\Omega_{\text{m},0}=0.24$, $\Omega_{\omega,0}=0.07$, i.e., 
$\Omega_{\Lambda,0}=0.84$, $\chi^2=53.6$.
For the flat model with small rotation
$\Omega_{\omega,0}=-0.01$, $\Omega_{\text{m},0}=0.36$, i.e., 
$\Omega_{\Lambda,0}=0.65$, $\chi^2=53.7$.
 
It again confirms our conclusion that on the base of pure statistical analysis
we could only select ``corridor'' of possible models. However, if we assume
that the Universe is flat $\Omega_{k,0}=0$, we obtain estimations for
$\Omega_{m,0}$, $\Omega_{\omega,0}$ what seems to be realistic.
 
One should note that we also could separately estimate the value of ${\cal M}$
for sample B and C. We obtain ${\cal M}=-3.42$ what is again in very
good agreement with result of Efstathiou {\it et al.\/} \cite{Efstathiou99}
(what for the ``combined'' sample obtain the value of ${\cal M}=-3.45$). 
However, if we use that value in our analysis it does not change 
significantly our results (value of $\chi^2$ does not change more then 
$1$ what is marginal effect for $\chi^2$ distribution for $53$ or $55$ 
degrees of freedom.
 
\begin{figure}
\includegraphics[width=0.75\textwidth]{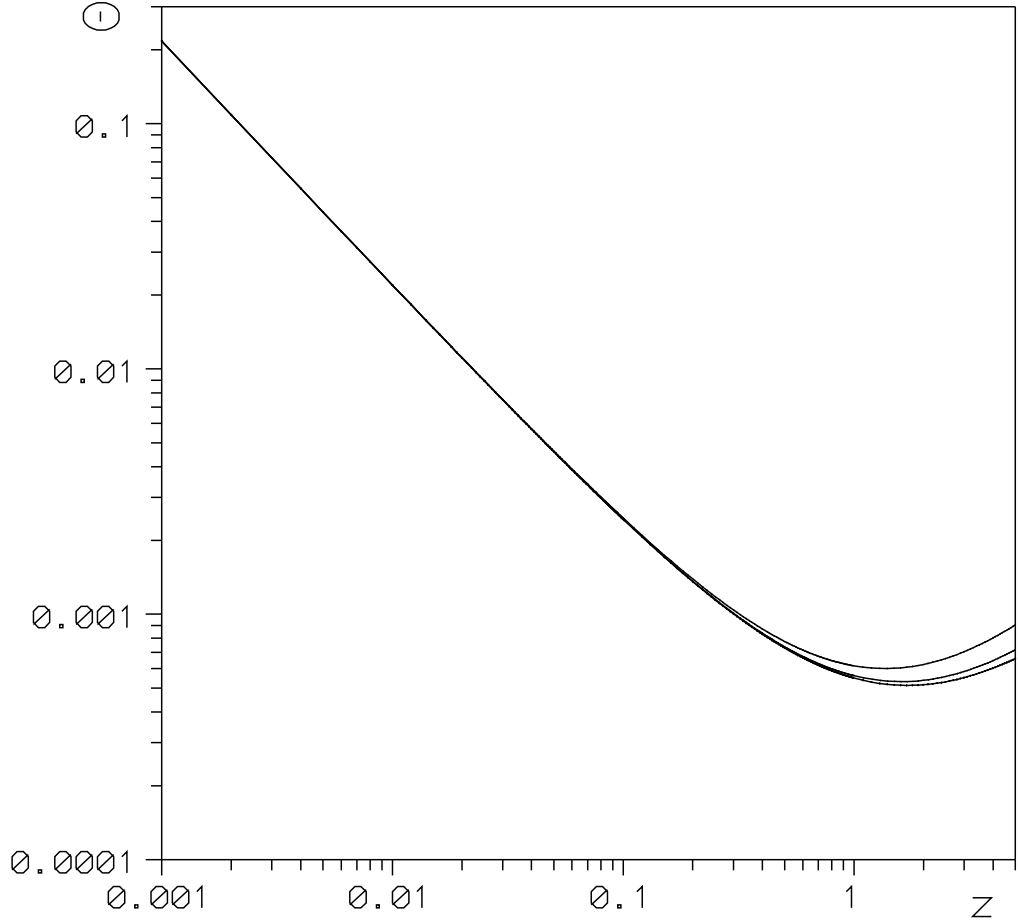}
\caption{The angular diameter $\Theta$ for the flat model with rotation
for $\Omega_{\text{m},0} = 0.3$ and $\Omega_{\text{r},0} = 0.1, 0, -0.02$ 
(top, middle, bottom). The minima for these cases are $1.364$, $1.605$, 
$1.707$, respectively. The rotation causes the minimum to move right 
(towards to higher $z$) and the minimum value of $\Theta$ decreases.}
\label{fig:4}
\end{figure}

We also analyse the influence of rotation for the age of the Universe. 
The results are presented on Fig.~\ref{fig:4}. If we assumed that 
$\Omega_{\text{m},0}=0.3$ and $H_{0}=65 \,\text{km}/\text{s}\, \text{Mpc}$
then small rotation $\Omega_{\omega,0}=-0.01$ increases the age of the 
Universe from $14.57\cdot 10^{10}\,\text{yr}$ to 
$15.17\cdot 10^{10}\,\text{yr}$.

Finally, let us study the angular diameter test for our universe.
The angular diameter of a galaxy is defined by
\begin{equation}
\label{angdiam}
\theta = \frac{d(z+1)^2}{d_{L}} ,
\end{equation}
where $d$ is a linear size of the galaxy. In a pure flat dust model
universe $\theta$ has the minimum value for $z_{\text{min}} = 5/4$.
It is particularly interesting to notice that for flat models with 
$\Omega_{\Lambda,0} \ne 0$ the dark radiation can increase the minimum 
value of $\theta$ toward the largest $z_{\text{min}}$ and smaller 
$\Theta_{\text{min}}$, while the ordinary radiation lowers this value.
 
\begin{figure}
\includegraphics[width=0.75\textwidth]{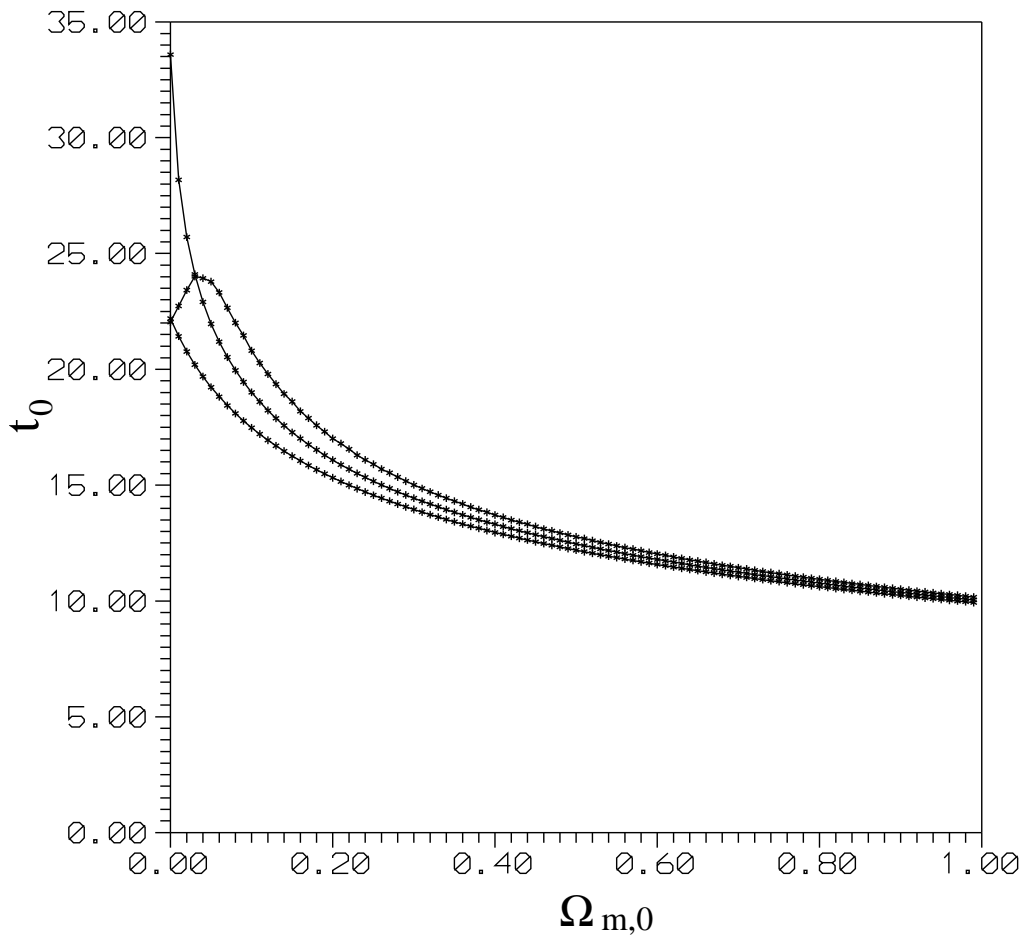}
\caption{The age of the universe $t_0$ in units of $10^{9}yr$
for the flat model with radiation $\Omega_{\text{r},0} = 0, 0.01$
(middle, bottom) and rotation $\Omega_{\omega,0}=-0.01$ (top).}
\label{fig:2a}
\end{figure}

We presents influence of rotation for the angular diameter $\Theta(z)$ as 
a function of redshift $z$. For the flat model with $\Omega_{m,0}=0.3$ as 
shown in Fig.~\ref{fig:2a}, the rotation causes the minimum to move right 
(higher $z$) and the minimum value of $\Theta(z)$ decreases. However, 
because there are small differences between predicted $\Theta(z)$ in all 
considered cases then verifying the observational test could be difficult.

\section{Conclusions}

We discuss the problem of universe acceleration driven by global rotation.
We demonstrate that the universe acceleration increase due to the
presence of global rotation effects, although the cosmological constant
is still required to explain SN Ia data. In this cosmology, the Friedmann
equation is modified by appearance of extra term which diminishes with
cosmic scale factor as $-a^{-4}$. Our model suggests limit for rotation
$\Omega_{\omega, 0}>-0.033$ (at one $\sigma$ level) if considered in the 
Newtonian model, however in the extended model with additional matter 
which scales like radiation  (not necessary $\Omega_{\text{r}, 0}<0$)
we obtain the more safely limit for rotation $\Omega_{\omega, 0}>-0.01$ 
(at one $\sigma$ level).

Our limit is weaker than that which can be obtained from BBN
($\Omega_{\omega, 0}=-1.23 \Omega_{\gamma,0}$) and CMB
($\Omega_{\omega, 0}=-0.41 \Omega_{\gamma,0}$) where present value of
$\Omega_{\gamma,0}$ is estimated as 
$\Omega_{\gamma,0} = 2.48h^{-2}\cdot 10^{-5}$ \cite{Vishwakarma02}.

We showed that, although the observational constraint from SN Ia allows
only a small contribution from `dark radiation' (however, when in the pure 
Newtonian model $\Omega_{\omega,0}<0$ a much wider range of negative values 
of $\Omega_{\omega,0}$ are allowed. We can find the strict analogy between 
the considered analysis of the observational constraints on the global 
rotation in the model and the search for observational constraints on dark 
radiation in brane cosmology. The corresponding term in brane cosmology 
scales just like radiation with a constant $\rho_0$ or both positive and 
negative $\rho_{\text{r},0}$ ($\rho_{\omega,0}$) are possible mathematically.
Dark radiation should strongly affect both the Big-Bang nucleosynthesis (BBN)
and the cosmic microwave background (CMB). Ichiki {\em et al.\/}
\cite{Ichiki02} used such observations to constrain both the magnitude
and the sign of dark radiation in the case when term $\rho^2$ coming
from the brane is negligible (it rapidly decays as $a^{-8}$ in the
early radiation dominated universe). Therefore, the presence of the term is
insignificant during the during the later nucleosynthesis. In such an
approximation we recover the considered model in which dark radiation
mimics radiation or rotation. Let us note negative contribution coming 
from the global rotation presence can reconcile the tension between the 
observed ${}^4\text{He}$ and $D$ abundance \cite{Ichiki02}. The application 
of these results gives also the possible constraints on global rotation term 
from BBN and from the power spectrum of CMB anisotropies.

We obtain the limit for $\Omega_{\omega,0}$ from BBN as $-7.21\cdot 10^{-5}$, 
while the limit from CMB is $-2.41\cdot 10^{-5}$. The present extragalactic 
data suggest $\omega_{0} = 6 \cdot 10^{-21} \text{rad}/\text{s}$ \cite{Li98}. 
This gives the strongest limit for $\Omega_{\omega,0}>-5.3\cdot 10^{-6}$.
Therefore, we can conclude that the present observational data of SN Ia 
give the weaker limit for rotation then obtained by other methods. 
However, let us note that the obtained limitations are constructed in 
independent manner.

\acknowledgments
M.S. was supported by KBN Grant No. 2 P03B 107 22.

%\bibliography{astronomy,rotacja}

\begin{thebibliography}{19}
\expandafter\ifx\csname natexlab\endcsname\relax\def\natexlab#1{#1}\fi
\expandafter\ifx\csname bibnamefont\endcsname\relax
  \def\bibnamefont#1{#1}\fi
\expandafter\ifx\csname bibfnamefont\endcsname\relax
  \def\bibfnamefont#1{#1}\fi
\expandafter\ifx\csname citenamefont\endcsname\relax
  \def\citenamefont#1{#1}\fi
\expandafter\ifx\csname url\endcsname\relax
  \def\url#1{\texttt{#1}}\fi
\expandafter\ifx\csname urlprefix\endcsname\relax\def\urlprefix{URL }\fi
\providecommand{\bibinfo}[2]{#2}
\providecommand{\eprint}[2][]{\url{#2}}

\bibitem[{\citenamefont{Godlowski et~al.}(2003)\citenamefont{Godlowski,
  Szydlowski, Flin, and Biernacka}}]{Godlowski03}
\bibinfo{author}{\bibfnamefont{W.}~\bibnamefont{Godlowski}},
  \bibinfo{author}{\bibfnamefont{M.}~\bibnamefont{Szydlowski}},
  \bibinfo{author}{\bibfnamefont{P.}~\bibnamefont{Flin}}, \bibnamefont{and}
  \bibinfo{author}{\bibfnamefont{M.}~\bibnamefont{Biernacka}},
  \bibinfo{journal}{Gen. Relat. Grav.} \textbf{\bibinfo{volume}{35}},
  \bibinfo{pages}{907} (\bibinfo{year}{2003}).

\bibitem[{\citenamefont{Li}(1998)}]{Li98}
\bibinfo{author}{\bibfnamefont{L.-X.} \bibnamefont{Li}}, \bibinfo{journal}{Gen.
  Relat. Grav.} \textbf{\bibinfo{volume}{30}}, \bibinfo{pages}{497}
  (\bibinfo{year}{1998}).

\bibitem[{\citenamefont{Collins and Hawking}(1973)}]{Collins73}
\bibinfo{author}{\bibfnamefont{C.~B.} \bibnamefont{Collins}} \bibnamefont{and}
  \bibinfo{author}{\bibfnamefont{S.~W.} \bibnamefont{Hawking}},
  \bibinfo{journal}{Mon. Not. Roy. Astr. Soc.} \textbf{\bibinfo{volume}{162}},
  \bibinfo{pages}{307} (\bibinfo{year}{1973}).

\bibitem[{\citenamefont{Hawking}(1974)}]{Hawking74}
\bibinfo{author}{\bibfnamefont{S.~W.} \bibnamefont{Hawking}}, in
  \emph{\bibinfo{booktitle}{Confrontation of Cosmological Theories with
  Observational Data}}, edited by \bibinfo{editor}{\bibfnamefont{M.~S.}
  \bibnamefont{Longair}} (\bibinfo{publisher}{Reidel Publ.},
  \bibinfo{address}{Dordrecht}, \bibinfo{year}{1974}), p. \bibinfo{pages}{283}.

\bibitem[{\citenamefont{Kristian and Sachs}(1966)}]{Kristian66}
\bibinfo{author}{\bibfnamefont{J.}~\bibnamefont{Kristian}} \bibnamefont{and}
  \bibinfo{author}{\bibfnamefont{R.~K.} \bibnamefont{Sachs}},
  \bibinfo{journal}{Astrophys. J.} \textbf{\bibinfo{volume}{143}},
  \bibinfo{pages}{379} (\bibinfo{year}{1966}).

\bibitem[{\citenamefont{Ellis}(1973)}]{Ellis73}
\bibinfo{author}{\bibfnamefont{G.~F.~R.} \bibnamefont{Ellis}}, in
  \emph{\bibinfo{booktitle}{Carg{\`e}se Lectures in Physics}}, edited by
  \bibinfo{editor}{\bibfnamefont{E.}~\bibnamefont{Schatzman}}
  (\bibinfo{publisher}{Gordon and Breach}, \bibinfo{address}{New York},
  \bibinfo{year}{1973}), vol.~\bibinfo{volume}{6}.

\bibitem[{\citenamefont{Ciufolini and Wheeler}(1995)}]{Ciufolini95}
\bibinfo{author}{\bibfnamefont{I.}~\bibnamefont{Ciufolini}} \bibnamefont{and}
  \bibinfo{author}{\bibfnamefont{J.~A.} \bibnamefont{Wheeler}},
  \emph{\bibinfo{title}{Gravitation and Inertia}}
  (\bibinfo{publisher}{Princeton University Press},
  \bibinfo{address}{Princeton}, \bibinfo{year}{1995}).

\bibitem[{\citenamefont{Senovilla et~al.}(1998)\citenamefont{Senovilla,
  Sopuerta, and Szekeres}}]{Senovilla98}
\bibinfo{author}{\bibfnamefont{J.~M.} \bibnamefont{Senovilla}},
  \bibinfo{author}{\bibfnamefont{C.}~\bibnamefont{Sopuerta}}, \bibnamefont{and}
  \bibinfo{author}{\bibfnamefont{P.}~\bibnamefont{Szekeres}},
  \bibinfo{journal}{Gen. Rel. Grav.} \textbf{\bibinfo{volume}{30}},
  \bibinfo{pages}{389} (\bibinfo{year}{1998}).

\bibitem[{\citenamefont{Heckmann and Sch{\"u}cking}(1959)}]{Heckmann59}
\bibinfo{author}{\bibfnamefont{O.}~\bibnamefont{Heckmann}} \bibnamefont{and}
  \bibinfo{author}{\bibfnamefont{E.}~\bibnamefont{Sch{\"u}cking}}, in
  \emph{\bibinfo{booktitle}{Handbuch der Physik}}, edited by
  \bibinfo{editor}{\bibfnamefont{S.}~\bibnamefont{Fl{\"u}gge}}
  (\bibinfo{publisher}{Springer-Verlag}, \bibinfo{address}{Berlin},
  \bibinfo{year}{1959}), vol. \bibinfo{volume}{LIII}, p. \bibinfo{pages}{489}.

\bibitem[{\citenamefont{Perlmutter et~al.}(1999)}]{Perlmutter99}
\bibinfo{author}{\bibfnamefont{S.}~\bibnamefont{Perlmutter}}
  \bibnamefont{et~al.}, \bibinfo{journal}{Astrophys. J.}
  \textbf{\bibinfo{volume}{517}}, \bibinfo{pages}{565} (\bibinfo{year}{1999}).

\bibitem[{\citenamefont{Riess et~al.}(1998)}]{Riess98}
\bibinfo{author}{\bibfnamefont{A.~G.} \bibnamefont{Riess}}
  \bibnamefont{et~al.}, \bibinfo{journal}{Astron. J.}
  \textbf{\bibinfo{volume}{116}}, \bibinfo{pages}{1009} (\bibinfo{year}{1998}).

\bibitem[{\citenamefont{Caldwell et~al.}(1998)\citenamefont{Caldwell, Dave, and
  Steinhardt}}]{Caldwell98}
\bibinfo{author}{\bibfnamefont{R.~R.} \bibnamefont{Caldwell}},
  \bibinfo{author}{\bibfnamefont{R.}~\bibnamefont{Dave}}, \bibnamefont{and}
  \bibinfo{author}{\bibfnamefont{P.}~\bibnamefont{Steinhardt}},
  \bibinfo{journal}{Phys. Rev. Lett.} \textbf{\bibinfo{volume}{80}},
  \bibinfo{pages}{1582} (\bibinfo{year}{1998}).

\bibitem[{\citenamefont{Weinberg}(1972)}]{Weinberg72}
\bibinfo{author}{\bibfnamefont{S.}~\bibnamefont{Weinberg}},
  \emph{\bibinfo{title}{Gravitation and Cosmology}}
  (\bibinfo{publisher}{Wiley}, \bibinfo{address}{New York},
  \bibinfo{year}{1972}).

\bibitem[{\citenamefont{Efstathiou et~al.}(1999)\citenamefont{Efstathiou,
  Bridle, Lasenby, Hobson, and Ellis}}]{Efstathiou99}
\bibinfo{author}{\bibfnamefont{G.}~\bibnamefont{Efstathiou}},
  \bibinfo{author}{\bibfnamefont{S.~L.} \bibnamefont{Bridle}},
  \bibinfo{author}{\bibfnamefont{A.~N.} \bibnamefont{Lasenby}},
  \bibinfo{author}{\bibfnamefont{M.~P.} \bibnamefont{Hobson}},
  \bibnamefont{and} \bibinfo{author}{\bibfnamefont{R.~S.} \bibnamefont{Ellis}},
  \bibinfo{journal}{Mon. Not. Roy. Astr. Soc.} \textbf{\bibinfo{volume}{303}},
  \bibinfo{pages}{L47} (\bibinfo{year}{1999}).

\bibitem[{\citenamefont{Vishwakarma}(2001)}]{Vishwakarma01}
\bibinfo{author}{\bibfnamefont{R.~G.} \bibnamefont{Vishwakarma}},
  \bibinfo{journal}{Gen. Relat. Grav.} \textbf{\bibinfo{volume}{33}},
  \bibinfo{pages}{1973} (\bibinfo{year}{2001}).

\bibitem[{\citenamefont{Peebles and Ratra}(2002)}]{Peebles02}
\bibinfo{author}{\bibfnamefont{P.~J.~E.} \bibnamefont{Peebles}}
  \bibnamefont{and} \bibinfo{author}{\bibfnamefont{B.}~\bibnamefont{Ratra}}
  (\bibinfo{year}{2002}), \eprint{astro-ph/0207347}.

\bibitem[{\citenamefont{Lahav}(2002)}]{Lahav02}
\bibinfo{author}{\bibfnamefont{O.}~\bibnamefont{Lahav}} (\bibinfo{year}{2002}),
  \eprint{astro-ph/0208297}.

\bibitem[{\citenamefont{Vishwakarma and Singh}(2002)}]{Vishwakarma02}
\bibinfo{author}{\bibfnamefont{R.~G.} \bibnamefont{Vishwakarma}}
  \bibnamefont{and} \bibinfo{author}{\bibfnamefont{P.}~\bibnamefont{Singh}}
  (\bibinfo{year}{2002}), \eprint{astro-ph/0211285}.

\bibitem[{\citenamefont{Ichiki et~al.}(2002)\citenamefont{Ichiki, Yahiro,
  Kajino, Orito, and Mathews}}]{Ichiki02}
\bibinfo{author}{\bibfnamefont{K.}~\bibnamefont{Ichiki}},
  \bibinfo{author}{\bibfnamefont{M.}~\bibnamefont{Yahiro}},
  \bibinfo{author}{\bibfnamefont{T.}~\bibnamefont{Kajino}},
  \bibinfo{author}{\bibfnamefont{M.}~\bibnamefont{Orito}}, \bibnamefont{and}
  \bibinfo{author}{\bibfnamefont{G.~J.} \bibnamefont{Mathews}}
  (\bibinfo{year}{2002}), \eprint{astro-ph/0203272}.

\end{thebibliography}
%\bibliographystyle{apsrev}

\end{document}